\pdfoutput=1
\documentclass[twocolumn,prl,amsmath,letterpaper,floatfix,showpacs]{revtex4}
\usepackage{graphicx}
\usepackage{amsmath}
\usepackage{amssymb}
\usepackage{color}
\usepackage{natbib}

\newcommand{\figref}[2]{Fig.~\ref{#1}~{\bf #2}}

\begin{document}

\title{Dynamics of molecular superrotors in external magnetic field}

\author{Aleksey Korobenko and Valery Milner}
\affiliation{Department of  Physics \& Astronomy, The University of British Columbia, Vancouver, Canada}

\begin{abstract}
We excite diatomic oxygen and nitrogen to high rotational states with an optical centrifuge and study their dynamics in external magnetic field. Ion imaging is employed to directly visualize, and follow in time, the rotation plane of molecular superrotors. The two different mechanisms of interaction between the magnetic field and the molecular angular momentum in paramagnetic oxygen and non-magnetic nitrogen lead to the qualitatively different behaviour. In nitrogen, we observe the precession of the molecular angular momentum around the field vector. In oxygen, strong spin-rotation coupling results in faster and richer dynamics, encompassing the splitting of the rotation plane in three separate components. As the centrifuged molecules evolve with no significant dispersion of the molecular wave function, the observed magnetic interaction presents an efficient mechanism for controlling the plane of molecular rotation.
\end{abstract}
\maketitle

Magnetic field is one of the most powerful tools for controlling atomic and molecular dynamics. Both translational and rotational degrees of freedom of molecules have been successfully manipulated via various types of magnetic interaction. Alignment of molecular axes has been accomplished by creating pendular states in paramagnetic molecules at low temperature \cite{Friedrich1992, Slenczka1994}. Gyroscopic precession along the direction of the applied magnetic field has been suggested for non-magnetic molecules in dispersionless ``cogwheel'' states \cite{Lapert2011, Yun2013}. Spin-rotational coupling has been recently utilized for converting unidirectional molecular rotation into axial alignment, manifested through magneto-rotational optical birefringence under ambient conditions \cite{Milner2015a, Floss2015}.

Recent progress in creating and detecting unidirectional rotation of molecules \cite{Fleischer2009, Kitano2009, Zhdanovich2011, Korobenko2014a, Mizuse2015, Lin2014} revived the interest in controlling molecular rotation with external magnetic fields. Magnetic effects could further broaden the controllability provided by tunable rotational excitation, including control of molecular collisions \cite{Tilford2004, Milner2014} and scattering of molecules at gas-solid interfaces \cite{Khodorkovsky2010}, molecular trajectories \cite{Gershnabel2010} and formation of gas vortices \cite{Steinitz2012}, optical \cite{Zahedpour2014, Milner2015a} and acoustic \cite{Schippers2011, Milner2015b} properties of a gas of rotating molecules.

With the invention of the optical centrifuge technique \cite{Karczmarek1999, Villeneuve2000}, a very flexible control over the degree of rotational excitation became available. The technique proved capable of producing synchronously rotating molecules with narrow rotational state distribution in a wide variety of species up to extremely high angular frequencies \cite{Yuan2011, Milner2015a}. However, the spatial orientation of the induced angular momentum is often defined by the propagation direction of the excitation laser beam, and is therefore not easy to manipulate.

Here we demonstrate how an applied magnetic field can be used to rotate the plane of molecular rotation. The described method is applicable to a wide variety of molecules and relies on the coupling between the rotation of a molecule and its magnetic moment. An applied magnetic field causes this moment to precess, ``dragging'' the axis of rotation with it. As discussed in this paper, the method is applicable to the molecules with magnetic moments of different nature, e.g. due to the rotation of the molecular frame or due to the electronic spin.

Consider a Hund's case b molecule possessing an electronic spin, such as molecular oxygen $^{16}O_2$ in the ground electronic state. Its electronic spin is strongly coupled to the molecular angular momentum $\mathbf{N}$. In an external magnetic field, sufficiently weak for not causing the spin to decouple from $\mathbf{N}$, the angular momentum will precess with frequency \cite{Floss2015}
\begin{equation}
\Omega=\mu_BgS_N\left\vert\mathbf{B}\right\vert\frac{1}{\hbar N},
\label{eq_ox}
\end{equation}
where $\mu_B$ is the Bohr's magneton, $g$ is the electron g-factor, $\mathbf{B}$ is the magnetic field strength, $S_N$ is the projection of an electronic spin on $\mathbf{N}$ and $\hbar$ is the reduced Plank's constant. The $1/N$ dependence in the above expression comes from the gyroscopic effect: at higher angular momenta, it becomes harder and harder for the same torque (determined by the product of the magnetic moment $\mu_BgS_N$ and the field strength $\mathbf{B}$) to change the rotational axis.

If the molecule does not have an electronic spin, its magnetic moment may stem from the electronic and nuclear currents (equal in magnitude and opposite in direction) encircling unequal areas as the molecule rotates. These leads to the net magnetic moment of the order of the nuclear magneton $\mu_N$ multiplied by the angular momentum. The magnetic precession in this case occurs with a frequency \cite{Yun2013}
\begin{equation}
\Omega=\mu_Ng_r\left\vert\mathbf{B}\right\vert\frac{1}{\hbar},
\label{eq_n2}
\end{equation}
where $g_r$ is the rotational g-factor. As the magnetic moment, and hence the torque, grows together with the angular momentum, no $N$-dependence is expected in this case. The precession frequency, however, is a few orders of magnitude lower compared to the magnetic molecules, due to the large difference between $\mu_N$ and $\mu_B$.

Using the technique of an optical centrifuge, we excited $O_2$ and $N_2$ molecules to high rotational states and observed the dynamics of their angular wave function in the magnetic field applied perpendicular to the rotation axis. By imaging the plane of molecular rotation with ion imaging, we followed its evolution in time and observed two different mechanisms of magnetic interaction, described above.
\begin{figure}[tb]
    \includegraphics[width=1\columnwidth]{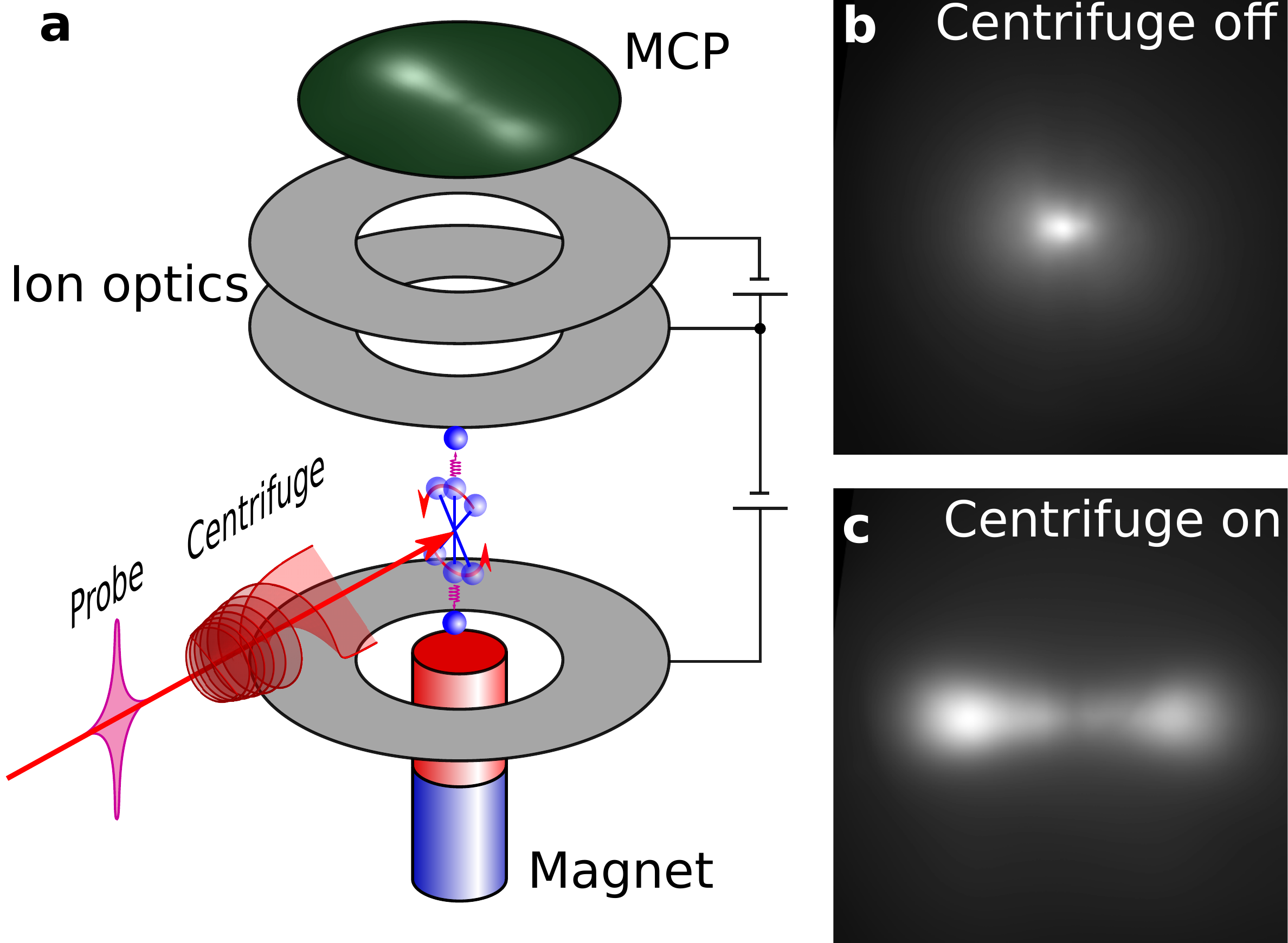}
    \caption{{\bf (a)} Experimental setup for detecting the plane of molecular rotation. The molecules are excited with a centrifuge pulse, then rotate freely in a magnetic field of a permanent magnet, and finally are ionized with a femtosecond probe pulse. The resulting atomic ions are extracted and focused with ion optics of a velocity map imaging (VMI) apparatus onto a microchannel plate detector (MCP) with a phosphorus screen. The images observed without {\bf (b)} and with {\bf (c)} the rotational excitation by a centrifuge pulse.}
    \label{fig_setup}
 \end{figure}

\section{Experimental setup}
Optical centrifuge pulses were prepared using the standard centrifuge shaper \cite{Karczmarek1999, Korobenko2014a}, starting from the output of a broadband Ti:Sapphire laser (Coherent, 35 nm FWHM bandwidth, centered at 800 nm). Part of the unshaped beam was used as a probe. Nitrogen or oxygen gas was supersonically expanded through a pulsed nozzle (Even-Lavie valve, 0.15 mm opening, 30 bar back pressure, 500 Hz repetition rate), which cooled most of the molecules down to the ground rotational state. The molecules were then excited with a centrifuge pulse to high (up to 10 THz) angular frequencies in between the plates of a time-of-flight (TOF) detector. A permanent rare-earth magnet, producing the vertical magnetic field of about 300 mT, was placed below the lower TOF plate (\figref{fig_setup}{a}).

After evolving in the applied magnetic field for a certain time period $t$, the molecules were exposed to an intense 40-femtosecond probe pulse, which lead to Coulomb explosion, i.e. multiple ionization and fragmentation into two atomic ions. In this process, the magnitude of the recoil momentum of a fragment ion is determined solely by the ionization channel, while its direction coincides with that of the internuclear axis at the moment of explosion. The fragment ions were accelerated towards the microchannel plate detector (MCP) with a phosphorus screen. Gating the voltage on the MCP with a nanosecond pulse generator allowed us to mass-select only doubly-ionized atomic ions, either O$^{2+}$ or N$^{2+}$.

For a given pulse energy ($\sim 10^{15}$ W/cm$^2$), the ions came primarily through a single ionization channel $X_2\longrightarrow X^{2+}+X^{+}+3e^{-}$. Due to the highly anisotropic ionization probability, molecules undergoing Coulomb explosion are mostly those aligned vertically along the linear polarization of the probe field. Hence, in the absence of the centrifuge, the observed image consisted of a single spot originated from the ions which had zero velocity component in the horizontal plane of the detector (\figref{fig_setup}{b}).

Rotational excitation by an optical centrifuge resulted in the transverse component of the ion recoil momentum due to the high angular velocity of the atomic nuclei at the moment of explosion. The energy of this transverse motion reached $\sim$0.86 eV for the highest accessible rotational states and was comparable to the Coulomb explosion energy of a few eV. As a result, the ions spread out horizontally in the plane of molecular rotation, as can be seen in \figref{fig_setup}{c}. The long axis of the observed ion image (horizontal axis in \figref{fig_setup}{c}) can therefore serve as a direct indication of the plane of molecular rotation. As expected, the degree of image elongation grew with the degree of rotational excitation.
\begin{figure*}[t!]
    \includegraphics[width=2\columnwidth]{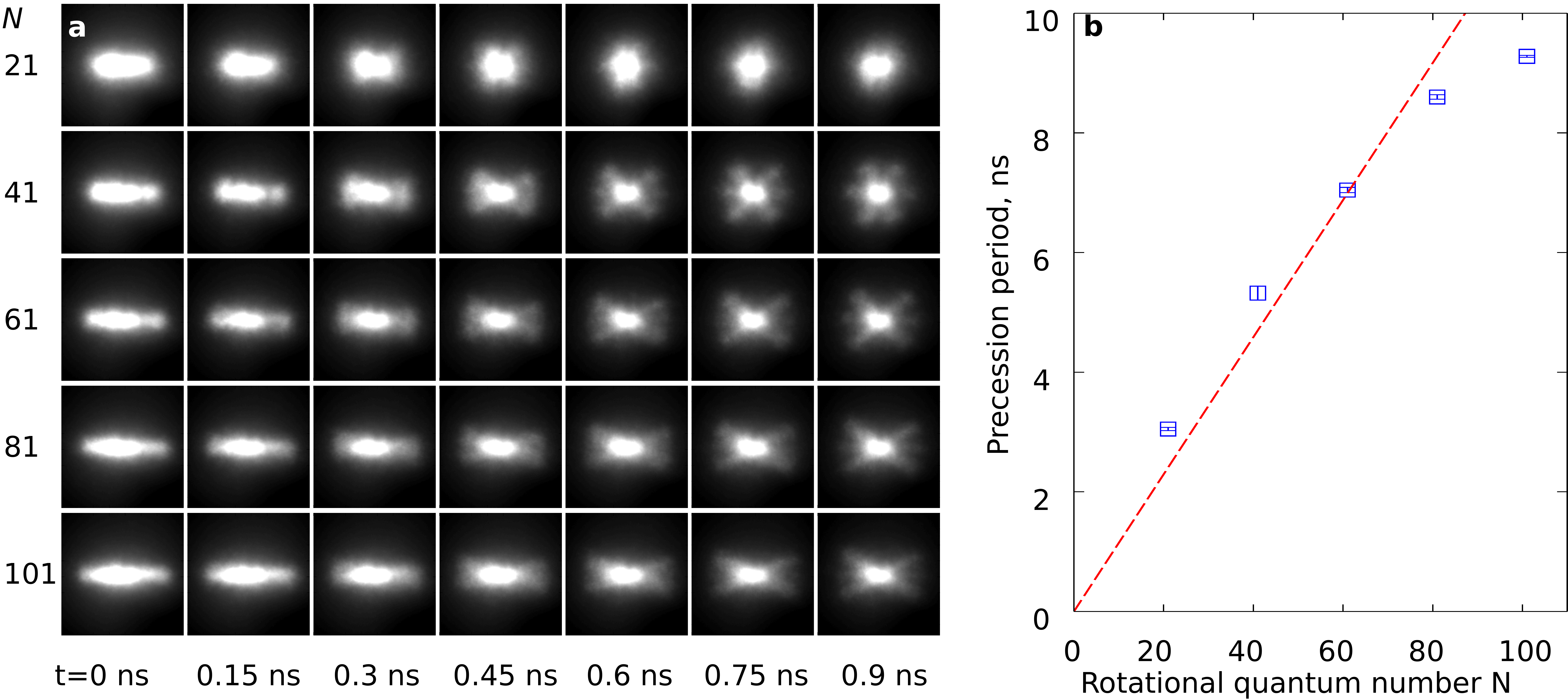}
    \caption{{\bf (a)} Ion images of oxygen superrotors evolving in external magnetic field. Different rows correspond to different degrees of rotational excitation ($N$ value on the left), whereas each column corresponds to the evolution time indicated at the bottom. The initial disk distribution splits into three disks precessing with different frequencies according to their spin projections. {\bf (b)} Experimentally determined precession period as a function of the rotational quantum number (blue squares) and the expected theoretical dependence (dashed red line).}
    \label{fig_oxygen}
\end{figure*}

\section{Results}
In the experiment on O$_{2}$, the delay between the centrifuge and probe pulses was controlled with a motorized delay line and changed from 0 to 0.9 ns. The final molecular angular momentum was varied by truncating the spectrum of an optical centrifuge as described in our earlier work \cite{Korobenko2014}. The observed ion images are shown in \figref{fig_oxygen}{a} for different degrees of rotational excitation (rows) and at different probe delays (columns).

For all observed angular momenta, as the pump-probe delay increases, the initial plane of molecular rotation splits into three: one which does not precess (horizontal axis in \figref{fig_oxygen}{a}) and two others, precessing in clockwise and counter-clockwise direction. The splitting into three planes, clearly visible at longer delays, results from the dependence of the precession frequency on the projection of the electronic spin on the rotational angular momentum $S_N$(see Eqn.(\ref{eq_ox})), which can take the values of 0 and $\pm1$.

It could be seen from the images, that the precession frequency decreases with increasing angular momentum, in agreement with Eq.\ref{eq_ox}. The quantitative comparison is shown in \figref{fig_oxygen}{b}, where the observed values (blue squares) are plotted together with the theoretical prediction (red dashed line) calculated for the magnetic field strength of $\left\vert\mathbf{B}\right\vert=316~\mathrm{mT}$, determined from the nitrogen experiments (as explained below). The experimental data is in good agreement with the predicted values, except of the higher angular momentum. We attribute this discrepancy to the poorer performance of the centrifuge at higher $N$ values. As the centrifuge field becomes weaker at higher angular velocities, some molecules are ``spilled over'' from the centrifuge prior to reaching its terminal rotational frequency.

\begin{figure}[t]
    \includegraphics[width=1\columnwidth]{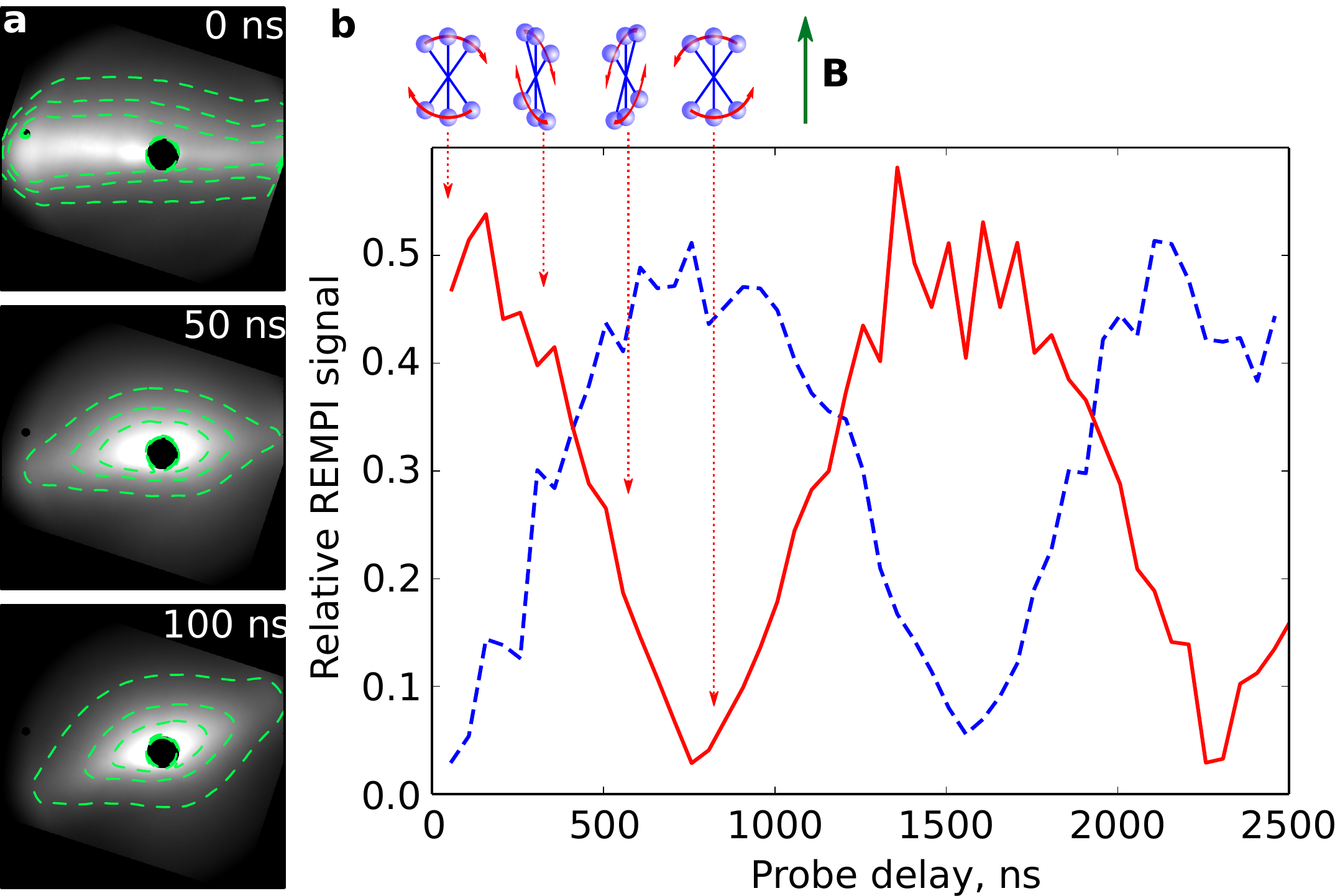}
    \caption{{\bf (a)} Ion images of nitrogen superrotors in external magnetic field. To improve the visibility of precession, an over-saturated central region is cut off and contour lines (green dashed) are added. {\bf (b)} Directional REMPI signal as a function of the pump-probe delay. Maximum (minimum) signal is detected when the molecules rotate in the opposite (same) direction with respect to the circularly polarized probe. Solid red and dashed blue lines correspond to the opposite initial directions of molecular rotation, i.e. opposite handedness of the centrifuge pulse.}
    \label{fig_nitrogen}
\end{figure}

We repeated the same experiment with molecular nitrogen. To observe the magnetic precession of this non-magnetic molecule, probe pulses were delayed for up to 100 ns. The resulting images, taken with the maximum available centrifuge excitation ($N\approx60$), are shown in \figref{fig_nitrogen}{a}. Although long time delays resulted in the reduced image quality, the precession is still clearly seen (notice dashed contour lines indicating counter-clockwise rotation).

To observe this precession at even longer delays, the directional resonance-enhanced multi-photon ionization (REMPI) approach was employed \cite{Kitano2009}. Centrifuged molecules were ionized with a dye laser, resonant with $a^1\Pi_g(v'=0, J'=31)\leftarrow\leftarrow X^1\Sigma^+_g(v''=0, J''=33)$ two-photon transition (285.813 nm), thus probing the population of the $N=33$ rotational level in the ground electronic state. The terminal rotational frequency of the centrifuge field was tuned to populate this state. In a centrifuged ensemble rotating primarily in one direction, the ionization probability is substantially different when the direction of rotation is the same as, or opposite to, the direction of the circular polarization of the probe light. Namely, when the direction of rotation coincides with that of the probe polarization, the two photon transition is forbidden by the conservation of the angular momentum projection. In the opposite case, i.e. for the molecules rotating against the probe polarization, the two-photon transition is allowed and results in strong ion yield.

Directional REMPI signal was measured as a function of the probe delay. At each time delay, the probe beam was moved downstream from the pump to account for the supersonic drift of the molecular cloud, which becomes relevant on these time scales. To compensate for the spread of the moving molecular cloud and the change in the beams overlap, each intensity was normalized to the signal obtained with vertical probe polarization at that delay time.

The results are shown in \figref{fig_nitrogen}{b}. When the direction of the centrifuge spinning was opposite to the probe polarization handedness, the measured signal was initially high (solid red line). However, in half the precession period, when the axis of molecular rotation turned by 180 degrees, the direction of rotation coincided with that of the probe polarization and the signal reached its minimum. These oscillations continued over the whole observed time range, with the plane of polarization completing a total of more than 1.5 full turns over 2500 ns, giving a precession period of 1550 ns. Given the rotational g-factor of N$_{2}$ ($-0.2681$ \cite{Cybulski1994}), this corresponds to the magnetic field strength of 316 mT, comparable with the expected value for our magnet. When the centrifuge direction was reversed (dashed blue line in \figref{fig_nitrogen}{b}), oscillations of similar period and amplitude, but shifted by $\pi $ radian, were observed in agreement with the expected behavior.

In conclusion, we have investigated the dynamics of molecular superrotors in external magnetic field, demonstrating the versatility of controlling the orientation of molecular angular momentum in both paramagnetic and non-magnetic molecules. The effect can be used to study high-$N$ molecular collisions with a gas target or a solid surface. Varying the magnetic field strength or the distance to the target, the orientation of the rotating molecules can be tuned from ``cartwheel'' to ``helicopter'' geometry. This approach may be further improved by employing a more complex field configuration, in which the magnetic field direction changes along the trajectory of the molecular jet from being initially parallel to the centrifuge beam to any desired final orientation. As the superrotors travel downstream, their angular momenta will adiabatically follow the direction of the applied magnetic field, regardless of the exact field strength, molecular angular momentum or time of travel.

We acknowledge many stimulating discussions with Ilya Sh. Averbukh. This research was supported by the grants from NSERC and CFI.

\providecommand{\newblock}{}

\end{document}